\documentclass{ws-rv9x6}
\usepackage{subfigure}   
\usepackage{ws-rv-thm}   
\usepackage{ws-rv-van}   

\usepackage{bm}
\usepackage{latexsym}
\usepackage{dcolumn}
\usepackage{amsmath,amsfonts,amssymb}
\usepackage{graphicx,epsfig}
\usepackage{verbatim}
\usepackage{xcolor}
\usepackage[caption=false]{subfig}
\usepackage{slashed}
\usepackage{etex}
\usepackage{float}
\usepackage{mathtools}
\usepackage{mathrsfs}
\usepackage{braket}
\usepackage{parskip}
\usepackage{setspace}
\usepackage[active]{srcltx}
\usepackage{wrapfig}
\usepackage[margin=1in]{geometry}
\usepackage{tikz}
\usetikzlibrary{arrows.meta}
\usepackage{graphicx}

\makeindex

\def\nn{\nonumber}

\def\l{\left}
\def\r{\right}
\def\DM{\mathrm{d}}
\def\lp{\ell_0}

\def\nn{\nonumber}

\def\l{\left}
\def\r{\right}
\def\DM{\mathrm{d}}
\def\lp{\ell_0}

\def \lp {L_0}

\def \lp {\ell_0}

\def \Rsq {\widetilde {\rm \bf Ric}(p;p_0)}

\def \Rsgp {{\rm \bf Ric}(p)}

\begin{document}

\chapter[
Quantum nature of gravity and spacetime 
\\ Fundamental insights from the Hoyle-Narlikar theory]
{Quantum nature of gravity and spacetime \\ \textit{Fundamental insights from the Hoyle-Narlikar theory}
\label{ra_ch1}}

\vspace{.5cm}

\author[Dawood Kothawala]{Dawood Kothawala\footnote{Author footnote.}}

\address{Centre for Strings, Gravity, and Cosmology, \\ Department of Physics,
IIT Madras, Chennai 600036, \\
dawood@iitm.ac.in\footnote{Affiliation footnote.}}

\begin{abstract}
The Hoyle-Narlikar action-at-a-distance formulation for gravity remains one of the most mathematically rigorous attempts to incorporate Mach principle into physics. A lesser known, but no less significant, is the fact that it is also one of the first in which two point functions play a key role in determining spacetime structure, and hence, are more fundamental than the metric. We argue that these ingredients have direct relevance for modern attempts to understand fundamental, emergent, and informational aspects of quantum spacetime. The non-local description of quantum spacetime, with Planck scale $\lp$ as a zero-point length, provides a mechanism through which the entire universe can inform local causal structure, \textit{à la} Mach, Hoyle, and Narlikar. In this chapter, we highlight certain key aspects of the HN theory and then discuss case studies from modern research that illustrate how the quantum spacetime might realize a version of the \textit{quantum gravitational Mach principle}.
\end{abstract}


\body

\tableofcontents

\vfill

{\footnotesize{
\it 
\noindent I think it was Hermann Bondi who once said that physics is such a consistent and connected logical structure that if one starts to investigate it at any point and if one pursues correctly every issue that branches away from one's starting point, in the outcome one will be led to understand the whole of physics. With Mach's principle, it seems something like that.
}

\noindent
- Fred Hoyle, in \textit{Mach's Principle: From Newton's Bucket to Quantum Gravity}, Eds. J. Barbour, H. Pfister, Birkhauser.
}

\section{JVN and ``action-at-a-distance"}\label{intro}

I remember going through the famous paper by Wheeler and Feynman (WF) on the absorber theory of radiation, rooted in action-at-a-distance formulation of electrodynamics,\cite{wf1945,wf1949} as an undergraduate student. To an undergrad studying this formulation alongside a project in quantum field theory, it was obvious that the source theory of Schwinger\cite{schwinger1966,schwinger1968} provided a natural setting for the quantum formulation of action at a distance electrodynamics. And once this is recognized, would not changing the source from \textit{charge} to \textit{mass} be the simplest next step leading towards gravitation? Technically, maybe, but conceptually, certainly not. Conceptually, such questions pertain to some of the most powerful and fundamental notions in physics, viz, inertia, gravitation, and space and time. It is with these ideas in mind that I got introduced to the extremely rich work of Hoyle and Narlikar, set on very precisely stated conceptual foundations and backed by rigorous mathematical derivations.\cite{hn1964,hnbook,hoyle1996}

Naturally, the first thing I did, soon after joining IUCAA as a research scholar in 2005, was to meet JVN. This meeting, and a course JVN offered an year later based on his monograph with Fred Hoyle \cite{hnbook}, led to some extremely fruitful insights into the subtleties of formulating field theories (classical or quantum) in terms of worldline actions. And since behaviour of a collection of worldlines - congruence of timelike curves - provides the most operationally basic way to characterize the curvature of spacetime, one is naturally led to ask what implications the action at a distance formalism (and its quantization) might have for understanding the quantum nature of spacetime itself. Here, then, was a direct hint about the interconnection between the formalism developed by Hoyle and Narlikar and the one I had just started working on - incorporating Planck length as zero-point length in the mesoscopic description of spacetime processes with my supervisor Prof Thanu Padmanabhan \cite{kothawala2013,kothawala2014,kothawala2015}. And so it happened that in my final year of PhD, having completed all the projects that comprised my thesis, I focused entirely on the question of new non-local observables one would need to describe spacetime near Planck scales. My basic focus was Synge's World function \cite{synge-gr,poisson2011} and that happened to be also the object that featured prominently in the mathematical details of Hoyle-Narlikar theory. I fixed a meeting with JVN again to discuss this. He listened with interest, and immediately asked if I could use my results to address the issue of self-energy of the electron, and whether the Planck scale in-built into the geometry couples to the Hubble scale coming from \textit{response of the universe}, something he had worked out with Hoyle.

Over the past decade, while working on deriving the various implications of the \textit{qmetric} (see later), never once have I lost sight of the fact that, coupled with the HN formalism, it can provide just the kind of non-locality via which the universe can inform the small scale structure of spacetime. In this contribution, I will first highlight certain key aspects of the HN model, and then present the reader with an explicit calculation using the qmetric that carry echoes of Mach principle and the HN theory.
\section{Two point functions as fundamental objects in the Hoyle-Narlikar theory}
    \begin{figure}[!htb]
    \centering 
    \includegraphics[width=1.25\textwidth]{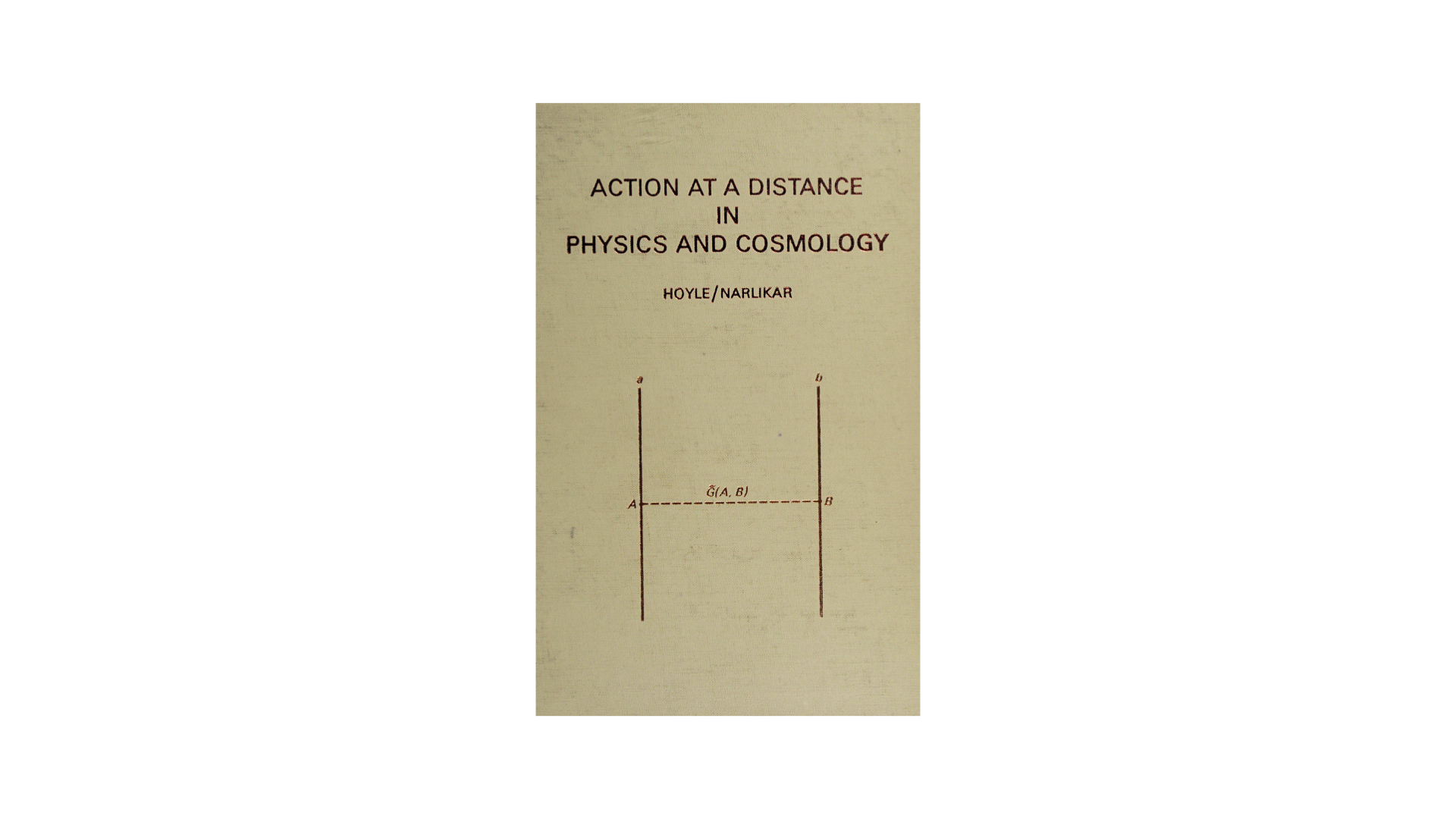} 
    \caption{The cover of the monograph by Hoyle and Narlikar very clearly highlights the central message of their theory; two point functions as the fundamental objects in physics.}
    \label{fig:my_image} 
\end{figure}
\subsection{Fokker action and Feynman-Wheeler absorber theory}
The very first attempt towards a relativistically invariant formulation of action at a distance electrodynamics was by Fokker,\cite{fokker1929} who proposed the following action for a system of charged particles: \cite{fokker1929,wf1945,wf1949}
\begin{eqnarray}
    I &=& \sum \limits_a I_{(a)}
    \\
    I_{(a)} &=& - \int m_{(a)} \DM a - q_{(a)} \int {A_i}(a) da^i
    \\
    {A_i}(a) &=& \sum \limits_b \overset{(b)}{A_i}(a)
    \\
    \overset{(b)}{A_i}(x) &=& q_{(b)} \int \delta(s_{x b}^2) \eta_{ij} db^j
\end{eqnarray}
We have written the action in a manner that will be easy to generalize to curved spacetime and gravitational interaction in a straightforward manner. In the above, $m_{(a)}$ and $q_{(a)}$ represent mass and charge of $a$th particle, $s_{x b}^2$ is the squared geodesic interval between events $x^i$ and $b^i$ (twice the so called Synge world function, $s^2<0$ for timelike separated points), and $\DM a$, $\DM b$ etc represent proper time intervals along trajectories of charges $a$, $b$ etc. The sum over all other charges $b$ is taken such that each pair $(a, b)$ in the original action counts only once; further restrictions on the sum come from demanding correct treatment of self action of a charge on itself. In any case, $\overset{(b)}{A_i}(a)$ represent the \textit{influence} of charge $b$ at/on $a$, and the sum of such influences from all the charges gives the effective vector potential felt by the $a$th charge. It can then be shown that the above action is equivalent to Maxwell equations and Lorentz force law for the charges, with the crucial difference that (1) advanced and retarded potentials are treated symmetrically, and (2) $\partial_i A^i=0$ if charges are neither created/destroyed. Feynman and Wheeler then showed that assuming the field produced by a charge, say $a$, were time symmetric (they took it as one half the sum of advanced and retarded fields), then, the sum of this field, plus the response field of all the other charges (``absorbers") in the future, can lead to results matching with conventional electrodynamics, including the radiation damping force on $a$.

There were, of course, problems with the Feynman-Wheeler formalism.\cite{hogarth1962} It treated the future and the past in an asymmetric manner, thereby introducing time asymmetry by hand. Besides, the collective response of the universe was captured via a refractive index, which was incorporated while evaluating effect of the source charge on the absorbers, but not considered in evaluating the effect of absorbers on the source charge.
\subsection{Hoyle and Narlikar's key insight}
The key insight by Hoyle and Narlikar, developed upon the important work by Hogarth, was to bring in the role of the time asymmetry induced by the evolving universe itself, and, most importantly, to generalize action at a distance electrodynamics to curved spacetime by shifting focus from the $\delta(s_{x b}^2)$ term above to the (scalar/vector) Green's functions of the d'Alembartian operator in curved spacetime.
\begin{eqnarray}
\Box_x G(x,y)
&=&
[-\bar{g}(x,y)]^{-1/2}\,
\delta^{(4)}(x,y)
\\
\Box_x G_{i j'}
+
R_{i}^{k}\,
G_{k j'}
&=&
[-\bar{g}(x,y)]^{-1/2}\,
\bar{g}_{i j'}\,
\delta^{(4)}(x,y)
\end{eqnarray}
where $\bar{g}_{ij'}$ is the parallel propagator, $\bar{g}$ its determinant, and primed indices refer to tangent space at $y$, while unprimed indices to tangent space at $x$. Hadamard's method of series expansions yields the following form for the above Green's functions \cite{hadamard1923, friedlander1975, dewittbrehme1960}
\begin{eqnarray}
G(x,y)
&=&
\frac{1}{4\pi}
\left[
\Delta^{1/2}\,
\delta\!\left(s_{xy}^{2}\right)
-
v\,\theta\!\left(-s_{xy}^{2}\right)
\right],
\label{eq:15}
\\
G_{ij'}
&=&
\frac{1}{4\pi}
\left[
g_{ij'}\,
\Delta^{1/2}\,
\delta\!\left(s_{xy}^{2}\right)
-
v_{ij'}\,
\theta\!\left(-s_{xy}^{2}\right)
\right].
\label{eq:16}
\end{eqnarray}
We shall not go into mathematical details of the various geometric objects appearing above, that have been studied extensively by de Witt and Brehme.\cite{dewittbrehme1960} The key quantities are the Synge world function and the van Vleck determinant $\Delta(x,y)={\rm det}\left(-\partial_i\partial_{j'}s^2/2\right)/{\bar g(x,y)}$. The quantities $v$ and $v_{ij'}$ are bi-tensors that arise solely from spacetime curvature and vanish in flat spacetime. They represent breakdown of Huygen's principle in curved spacetimes,\cite{dewittbrehme1960} and \textit{plays a crucial role in the problem of self action of charge on itself}, because of the $\theta(-s_{xy}^2)$ term.

The Hoyle-Narlikar generalization of Fokker action was to replace, in the latter,
\begin{eqnarray}
\overset{(b)}{A_i}(a) = q_{(b)} \int G_{i j'} db^{j'}
\end{eqnarray}
with the integral along the trajectory of $b$. Using
the important identity $\nabla^i G_{i j'} = - \nabla_{j'} G$, which relates objects that are vectors at $y$ and scalars at $x$, it is easy to see that the above vector field is divergenceless if there is no creation or destruction of charges in the infinite past/future.
\subsection{Generalisation to arbitrary field theories by JVN}
In subsequent papers, JVN established a concrete action at a distance formulation for all conventional field theories,\cite{narlikar1970} which goes as follows: Consider a field $\Phi$ that can represent a tensor field of arbitrary rank $N$, and a set of particles interacting with this field. The conventional action is given by
\begin{eqnarray}
    I[\Phi, a] = \int \mathscr{L}[\Phi] \sqrt{-g} \; \DM^4x + \sum \limits_a \int I\left[\Phi, a\right] \DM a
\end{eqnarray}
where the lagrangian $\mathscr{L}$ is a bi-linear in the field and its first derivatives, and $I$ represents a linear interaction of the form $\alpha_0 \, \Phi \circ \xi$ with coupling strength $\alpha_0$, $\xi$ a rank $N$ tensor field built entirely from source trajectories, and $\circ$ representing contraction over all indices. JVN then showed that the action at a distance theory that yields the same equations of motion is given by
\begin{eqnarray}
    I_{HN}[\Phi, a] \propto \alpha_0^2
    \int \int \overset{(a)}{\xi} \circ \mathscr{G} \circ \overset{(b)}{\xi}
    \DM a \; \DM b
\end{eqnarray}
where $\mathscr{G}$ is the Greens function of $\Phi$ field, and hence a rank $N$ bi-tensor at each point of its arguments.

Coming finally to gravity, Hoyle and Narlikar took a route guided by Mach principle, and considered the \textit{mass field}, or inertia, associated with particles as more fundamental \cite{hn1964}. The action proposed is
\begin{eqnarray}
    I \propto \sum \limits_a\sum\limits_b\int\int \limits \mathscr{G} \left[a, b \right] da\,db
\end{eqnarray}
where $\mathscr{G} \left[a, b \right]$ is proposed to be the (symmetric) Green's function of the conformally invariant wave operator. The variation of the above action with respect to the metric tensor gives a complicated equation that generalizes Einstein equations with additional terms.
\section{Relevance of HN ideas for quantum physics in/of curved spacetime}
In this section, we consider two important case studies with the intent to highlight how the basic mathematical and conceptual formulation of HN theory might be relevant for modern research on the interface of gravity and quantum field theory.
\subsection{Quantum entanglement in curved spacetime}
Perhaps the most natural generalization of the set-up considered by Feynman-Wheeler and Hoyle-Narlikar, to include quantum effects, is to start by replacing the sources (charge, mass, etc.) with quantum probes. The simplest model for these are the so-called Unruh-DeWitt (UDW) detectors,\cite{unruh1976,dewitt1979} two level probes (with energy gap $\omega$) that couple linearly to a background quantum scalar field $\Phi(x)$, and probe the vacuum structure of these fields. Two such UDW detectors, say $A$ and $B$, can be used to study quantum entanglement in curved spacetimes. It can be shown that the field mediated entanglement between the detectors, measured via the so called \textit{Negativity} $\mathcal{N}$,\cite{vidal2002,reznik2003,hari2024} is to lowest order given by
\begin{eqnarray}\label{eq:neg}
    \mathcal{N}(\rho_{\text{AB}}) &=& \frac{1}{2}\l[\sqrt{(\mathcal{I}_{\text{A}}-\mathcal{I}_{\text{B}})^2+4\,|\mathcal{I}_{\mathcal{E}}|^2} - (\mathcal{I}_{\text{A}}+\mathcal{I}_{\text{B}})\r]
    ~,
\end{eqnarray}
where
\begin{subequations}\label{eq:Ij-Ie-general}
\begin{eqnarray}\label{eq:Ij-general}
\mathcal{I}_{\text{A}} &=&
\int_{-\infty}^{\infty} d\tau^\prime \int_{-\infty}^{\infty}d\tau \;
\chi(\tau) \chi(\tau^{\prime})
e^{i \omega (\tau - \tau^{\prime})}
G_{\textsf{W}}(x_{\text{A}}(\tau^{\prime}), x_{\text{A}}(\tau))
\nonumber \\
i \, \mathcal{I}_{\varepsilon} &=& \int_{-\infty}^{\infty} ds \int_{-\infty}^{\infty}d\tau \; \chi(\tau) \chi(s)
e^{i \omega (\tau + s)}
G_{\textsf{F}}(x_{\text{B}}(s), x_{\text{A}}(\tau))~
\nonumber \label{eq:Ie-general}
\end{eqnarray}
\end{subequations}
where, $G_{\textsf{W}}(x^{\prime},x)$ and $G_\textsf{F}(x_2,x_1)$ respectively denote the Wightmann function and the Feynman propagators.\cite{birrell1982,dewitt1975} For gapless detectors ($\omega=0$) the above integrals are of the type appearing in the HN formalism, the only difference being that instead of the time symmetric combination of advanced and retarded Green's function, they depend on QFT propagators that carry information about the quantum state of the field. Considering the above integrals \textit{a la} Hoyle and Narlikar would give important insights into how the response of the universe affects quantum entanglement. It might also shed light on the quantum nature of the HN theory. To carry the above program forward, one would need to introduce a large number of detectors and then evaluate the effect on one of them of summing over all the other detectors. 
\subsection{Zero-point length and the Quantum spacetime}
Can the entire universe conspire to determine the local physical observables? This question, first raised by Mach,\cite{mach1893} have had connotations that go well beyond physical considerations, into the domain of philosophy. This has perhaps led to a mix up between the mathematical realization of Mach's ideas from the very vague statements about it by Mach himself. It was the HN formalism that gave the Machian ideas about physics a firm mathematical basis. Unfortunately, the mathematical non-triviality of this formalism has gotten lost in the poorly motivated attempts to apply it to cosmology.

Going a step ahead, one might ask: Can the entire universe conspire to determine the local structure of spacetime? Such a deeper, quantum gravitational, version of Mach principle would almost certainly have elements of HN theory in it, because of the way the latter is formulated. Distilling the key essence of Mach principle and HN theory, a connection between local spacetime structures and the universe at large can be expected to arise from a fundamental non-locality in the description of spacetime, coupled to its structure at smallest scales. Our aim in this section would be discuss a formalism that does this, and given a simple calculation within this formalism that demonstrates the claim.
\subsubsection{Machian threads to weave a quantum spacetime}
Several arguments from quantum gravity indeed suggest that quantum spacetime without a preferred frame would be necessarily non-local. The mathematical tools most appropriate for describing such a non-local spacetime can not be based on standard local tensorial objects such as $g_{ab}(x), R_{abcd}(x),$ etc., but rather on non-local observables that characterize measurements and observations. With such non-local observables in hand, one can try to re-construct the {\it quantum} spacetime geometry, at least in the mesoscopic domain, with some physically motivated input from semi-classical or quantum gravity. Ideally, one would like to have an input that arises from some broad principles of general relativity and quantum field theory, and is likely to survive any sieve that selects the final, correct framework of quantum gravity.

To summarize the above discussion: The mesoscopic domain of a non-local spacetime requires a specification, or choice, of (i) {\it Non-local classical observables}, and
(ii) a {\it physical input} from semi-classical and/or quantum gravity that can be recast in terms of these observables. To specify these in the simplest possible way, begin by considering two spacetime events $p_0$ and $p$, and ask: what are the most primitive geometric variables that one can associate with these points? If the events are in a geodesically convex normal neighborhood of each other, so that there is a unique geodesic $\cal C$ connecting them, the simplest choice for the inputs stated in the previous paragraph would be: (i) \underline{\textit{Non-local classical variables}}: {\it Synge's World function} \cite{synge-gr}, $\Omega(p_0,p)$ which equals one half the square of geodesic distance along {$\cal C$}, and the {\it van-Vleck determinant} $\Delta(p_0,p)$ that governs focussing of geodesics emanating from $p_0$ or $p$, and (ii) \underline{\textit{Physical input}}: There exists a lower bound on measurement of spacetime intervals. Using (i) alone, one can deduce a general structure for an ``{\it effective metric}" $q_{ab}$ which will mathematically be a non-local bi-tensor, since it is constructed from non-local objects. The precise form of $q_{ab}$ is then fixed by imposing (ii). The key role is now played by the Synge world function $\Omega(x, x')$, from which the geometry of spacetime must be deduced. This is indeed possible due to the well known properties of the World function and coincidence limits of its various higher derivatives.\cite{synge-gr} For instance, $\lim \limits_{x' \to x} \nabla_a \nabla_{b} \Omega(x, x') = g_{ab}(x)$, and hence the metric can indeed be re-constructed from $\Omega(x, x')$. Similarly, higher derivatives of $\Omega(x, x')$ carry information about spacetime curvature etc.

From the above discussion, it immediately follows that the most general form for $q_{ab}$ would be
\begin{eqnarray}
q_{ab}(p; p_0, \lp) &=& \textsf{A}\l[\xi\r] g_{ab} (p) - \epsilon \textsf{B}\l[\xi\r] t_a(p) t_b(p)
\end{eqnarray}
where $\xi = {\lp^2}/{\Omega(p_0, p)}$ and $\epsilon=g^{ab} t_a t_b=\pm 1$, and $\textsf{A}>0$ and $\textsf{B}>0$ are dimensionless functions (which, for now, we shall leave undetermined), with the limits: $\textsf{A}[0]=1, \textsf{B}[0]=0$. This ensures that at large scales, $q_{ab} \to g_{ab}$. We will also frequently use $q^a(p)=g^{ab}(p) t_b(p)$. It should be noted that the object $q_{ab}(p; p_0, \lp)$ is a {\it bi-tensor}: it behaves as a scalar with respect to coordinate transformations at $p_0$, and as a second-rank tensor with respect to coordinate transformations at $p$. This will be crucial for the averaging we propose below.

The above form for the quantum metric is quite general. Further restrictions, motivated by semi-classical arguments and/or some results from quantum gravity, can be used to fix the form of $\textsf{A}$ and $\textsf{B}$. In fact, the existence of a zero-point length—the requirement that spacetime intervals be bounded below—along with Lorentz invariance, are strong enough to fix the form of these functions. This has been discussed in general elsewhere.\cite{kothawala2013, kothawala2014, kothawala2015} Remarkably, here we will not need explicit form of these structure functions; our key insight will follow from their general behavior.
\subsubsection{Averaging the quantum spacetime}
We are now ready to introduce our key novel idea of this section, inspired by the ideas of Wheeler, Feynman, Hoyle, and Narlikar: treat $q_{ab}(p; p_0, \lp)$ as an effective metric at $p$ \textit{influenced} by $p_0$, and compute local quantities by averaging over all $p_0$ in the past light cone of $p$, with the averaging measure being the volume measure corresponding to $g_{ab}$. Averaging of tensors in curved spacetime is a notoriously subtle issue, since naive notions of it violate general covariance and a covariant notion is mathematically extremely challenging to handle.\cite{zalaletdinov1992} Our prescription above does not suffer from any of these difficulties since the qmetric is a scalar at $p_0$.
    \begin{figure}[!htb]
    \centering 
    \includegraphics[width=.5\textwidth]{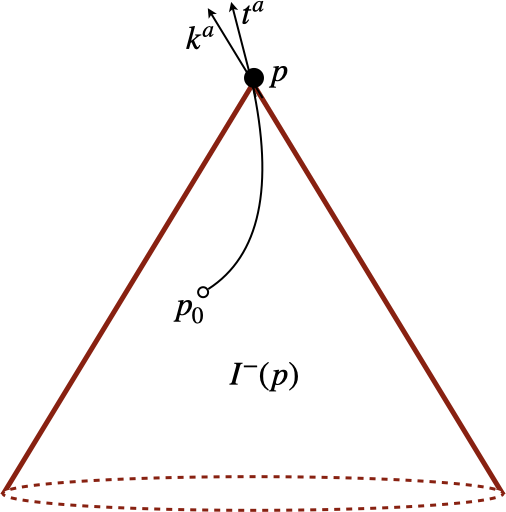} 
    \caption{Averaging the qmetric over the past light cone.}
    \label{fig:avg_lightcone} 
\end{figure}
We will apply the above averaging to the light cone structure of quantum spacetime. Let us assume that spacetime remains Lorentzian at small scales. It is then easy to see that $q_{ab} q^a q^b<0 \implies \textsf{A}\l[ \xi \r] -\textsf{B}\l[ \xi \r] >0$. The equations for null cones of $q_{ab}$ at $p$ becomes
\begin{eqnarray}
0 = \textsf{A} \l[ \xi \r] g_{ab} k^a k^b + \textsf{B}\l[ \xi \r] \l( k^a t_a \r)^2
\end{eqnarray}
from which it immediately follows that the vector $k^a$ is timelike with respect to $g_{ab}$, and that one may set $g_{ab} k^a k^b=-1$. Since $t_a$ by assumption is timelike, we can find the relative boost, $\gamma_{\rm rel} = -t_a k^a$, between $t_a$
and $k^a$. This immediately yields the {\it speed of light} as
\begin{eqnarray}
v_{\rm rel}^2 &=& 1 - \gamma_{\rm rel}^{-2}
\nn \\
&=& 1 - \frac{\textsf{B}\l[ \xi \r] }{\textsf{A}\l[ \xi \r] }
\end{eqnarray}
which, from our discussion above, satisfies $v_{\rm rel} < 1$ (this is also expected from the fact that $k^a$ is timelike).

Our averaging prescription stated above, applied to this ``speed of light" $v_{\rm rel}$, then gives the effective speed of light due to quantum corrections. Mathematically,
\[
c_{\rm eff} = \langle v_{\rm rel} \rangle
=
\frac{
\int \limits_{p_0 \in I^-(p)} v_{\rm rel} \sqrt{-g} \, d^4x
}{
\int \limits_{p_0 \in I^-(p)} \sqrt{-g} \, d^4x
}
\]

Explicit evaluation of the above integral requires structure of the metric in geodesic coordinates, as well as some information about how zero-point length is incorporated in $q_{ab}$. Let $S(\lambda^2)$ be the World function corresponding to $q_{ab}$, where $\lambda^2=|\Omega|$. We can write its most general form as
$S(\lambda^2)=\lambda^2 \left(1+K(\lambda^2/\ell_0^2)\right)$, where $K(x)$ is such that, it diverges as $1/x$ as $x \to 0$ and vanishes as $x \to \infty$. The detailed form of $K(x)$ must come from quantum gravity, but we will not need it here. All we need are its above limits, that ensure $S(0)=\ell_0^2$ - the condition for zero-point length. Imposing further the physically reasonable conditions $S>0, S'>0$, it can be shown that
\begin{eqnarray}
0 < 1+\frac{x K'(x)}{1 + K(x)} < 1
\end{eqnarray}

On the other hand, it can be shown that
\begin{eqnarray}
c_{\rm eff} = \frac{1}{
\int_0^{\xi_0} d\mu_\xi
}
\int_0^{\xi_0} \left[ 1+\frac{\xi K'(\xi)}{1 + K(\xi)} \right] r^{-1/3} d\mu_\xi
\end{eqnarray}
where $\xi_0=\lambda_0^2/\ell_0^2$, with $\lambda_0$ being the geodesic length of the averaging region. It is this scale that brings in the size of the universe into play. The measure $d\mu_\xi$ is known in terms of $g_{ab}$, while the precise form of $r$ can be determined in terms of the van Vleck determinant. If $r>1$ (it can be shown that this follows from geometric version of dominant energy condition), we arrive (upon restoring $c$) at the result:
\begin{eqnarray}
c_{\rm eff} = c \left[ 1 - \mathscr{J}(\ell_0^2/\tau^2) \right]
\end{eqnarray}
with $\mathscr{J}$ being essentially the factor $-{\xi K'(\xi)}/(1 + K(\xi))$, expressed for convenience as a function of $1/\xi$, so that
$\lim \limits_{x \to 0} \mathscr{J}(x)=0$ and $\lim \limits_{x \to \infty} \mathscr{J}(x)=1$, and we have replaced $\lambda_0$ with $\tau$ to emphasize its physical relevance as time scale of averaging.
\subsubsection{Can there be observational signatures of this? Non-local relics of a quantum spacetime}
Any signal which is proposed to have originated in the very early universe would inevitably carry an imprint of quantum effects of spacetime. However, it seems more likely that observable effects, such as those induced by $c_{\rm eff} \neq c$, should come from $\dot c_{\rm eff}$. We must caution, however, that quantifying any observable effects would require independent analysis along the lines given here for the corresponding observable. The algorithm for averaging prescribed here would then be a blueprint for such an analysis. As a final comment, we must point out that, if the age of the universe is bounded above $\tau^2<\Lambda$, then our result relates the quantity $\Lambda \ell_0^2$ with $c_{\rm eff}$, a connection worth exploring in depth for obvious reasons.
\subsection{Non-local gravitational action}
From the qmetric yield one can obtain a closed form expression for the Ricci bi-scalar $\Rsq$, and we must ask what a more fundamental action built from this object might look like, without resorting to any expansion in any of the parameters. Such an action must be evidently non-local, and one can only speculate as to its structure. Doing so will take us one step beyond the classical action proposed by Hoyle and Narlikar, due to its dependence on $\lp$. Towards this end, consider the following construction \cite{kothawala2023-essay}: Consider any event $p$ at which you construct the qmetric anchored on an event $p_0$, and compute the integral of
$\Rsq$ over all $p_0 \in I^-(p)$ - the causal past of $p$ - with measure $\DM \mu = \DM v(p_0)/v^-(p)$, where $\DM v(p_0)$ is a local volume measure, and $v^-(p)$ is the volume of $I^-(p)$ with respect to this measure, and ensures the normalization $\int_{p_0 \in I^-(p)} \DM \mu(p,p_0) = 1$. Integrate the final result over all $p$. Mathematically,
\begin{eqnarray}
{\rm Action} = \int \limits_{{\rm all} ~ p} \DM v(p) \int \limits_{p_0 \in I^-(p)} \Rsq \; \DM \mu(p,p_0)
\end{eqnarray}

Note that, when $\lp=0$,
$\Rsq = \Rsgp$, and we recover the standard Einstein-Hilbert action. On the other hand, keeping $\lp$ non-zero, the contribution from events $p_0$ ``close" to $p$ would give some weighted integral of the \textit{entropy functional} \cite{padmanabhan2010entropy} to which the qmetric Ricci scalar reduces in this limit. At intermediate scales, the above action would generically have a complicated structure, and a complete understanding of the corresponding variational principle would require much greater effort. It is here that we expect the HN results to provide the right tools and insights to proceed.
\section{Legacy of JVN}
While my academic interactions with JVN largely revolved around the mathematical details of action-at-a-distance formulation of gravity, I also had the opportunity to discuss with him his ideas on \textit{panspermia}, including the high-altitude balloon experiments supported by ISRO that he helped initiate in search for observational support of this hypothesis.\cite{wainwright2003} A senior colleague, Gaurang Mahajan, and myself prepared a National Science Day poster describing the motivation and key outcomes of this project. Unsurprisingly, it drew considerable interest from the general public throughout the day. As the day progressed, JVN himself came down from his office to see how the interactions were going on. While the school children listened attentively, it was the accompanying parents who immediately surrounded him, many greeting him with the reverence traditionally reserved for a guru in the Indian gurukul tradition. It is difficult to think of another Indian scientist who commanded such spontaneous affection and recognition from the general public.

Prof. Naresh Dadhich, who sadly passed away last year, recounted a similar episode from Pune during a public lecture by Sir Roger Penrose on black holes. According to his recollection, the anticipation that JVN would also be present drew such an overwhelming crowd that the gathering became difficult to manage. Order was restored only after JVN himself agreed to address the audience briefly before Penrose's lecture commenced. The episode speaks volumes about the extraordinary place JVN occupied—not only within the scientific community but also in the imagination of the wider public.

JVN's immense contributions to science—especially to Indian science—through his pioneering research, technical and popular science books and articles, science fiction stories and novels, a movie \textit{Dhoomketu} (``The Comet") challenging blind faith and superstition, a television series inspired by Carl Sagan's \textit{Cosmos}, and, above all, the creation of IUCAA itself, constitute a legacy of extraordinary breadth and significance. His legacy should inspire present and future generations of scientists to pursue their research with uncompromising integrity, intellectual courage, and maturity to distill scientific insights from fashion and prejudices that inevitably exist in science.
%
%


\bibliographystyle{ws-rv-van}
\bibliography{references}

\begin{thebibliography}{30}
\providecommand{\natexlab}[1]{#1}
\providecommand{\url}[1]{\texttt{#1}}
\expandafter\ifx\csname urlstyle\endcsname\relax
  \providecommand{\doi}[1]{doi: #1}\else
  \providecommand{\doi}{doi: \begingroup \urlstyle{rm}\Url}\fi

\bibitem{wf1945}
J.~A. Wheeler and R.~P. Feynman, Interaction with the absorber as the mechanism of radiation, \emph{Reviews of Modern Physics}. {\bf 17}, \penalty0 157--181  (1945).

\bibitem{wf1949}
J.~A. Wheeler and R.~P. Feynman, Classical electrodynamics in terms of direct interparticle action, \emph{Reviews of Modern Physics}. {\bf 21}, \penalty0 425--433  (1949).

\bibitem{schwinger1966}
J.~Schwinger, Particles and sources, \emph{Physical Review}. {\bf 152}\penalty0 (4), \penalty0 1219--1226  (Dec., 1966).
\newblock \doi{10.1103/PhysRev.152.1219}.

\bibitem{schwinger1968}
J.~Schwinger.
\newblock Theory of sources.
\newblock In \emph{Contemporary Physics: Trieste Symposium 1968}, vol.~2, pp. 59--110. International Atomic Energy Agency, Vienna  (1969).
\newblock Based on lectures delivered at the 1968 International Centre for Theoretical Physics (ICTP) Trieste Symposium.

\bibitem{hn1964}
F.~Hoyle and J.~V. Narlikar, A new theory of gravitation, \emph{Proceedings of the Royal Society A}. {\bf 282}, \penalty0 191--207  (1964).

\bibitem{hnbook}
F.~Hoyle and J.~V. Narlikar, \emph{Action at a Distance in Physics and Cosmology}. W. H. Freeman  (1974).

\bibitem{hoyle1996}
F.~Hoyle and J.~V. Narlikar, \emph{Lectures on Cosmology and Action-at-a-Distance Electrodynamics}. vol.~1, \emph{World Scientific Series in Astronomy and Astrophysics}, World Scientific, Singapore  (1996).
\newblock ISBN 9789810225582.

\bibitem{kothawala2013}
D.~Kothawala, Minimal length and small scale structure of spacetime, \emph{Physical Review D}. {\bf 88}, \penalty0 104029  (2013).
\newblock \doi{10.1103/PhysRevD.88.104029}.

\bibitem{kothawala2014}
D.~Kothawala and T.~Padmanabhan, Entropy density of spacetime as a relic from quantum gravity, \emph{Physical Review D}. {\bf 90}\penalty0 (12), \penalty0 124060  (2014).
\newblock \doi{10.1103/PhysRevD.90.124060}.

\bibitem{kothawala2015}
D.~J. Stargen and D.~Kothawala, Small scale structure of spacetime: The van vleck determinant and equi-geodesic surfaces, \emph{Physical Review D}. {\bf 92}\penalty0 (2), \penalty0 024046  (2015).
\newblock \doi{10.1103/PhysRevD.92.024046}.

\bibitem{synge-gr}
J.~L. Synge, \emph{Relativity: The General Theory}. North-Holland  (1960).

\bibitem{poisson2011}
E.~Poisson, A.~Pound, and I.~Vega, The motion of point particles in curved spacetime, \emph{Living Reviews in Relativity}. {\bf 14}, \penalty0 7  (2011).
\newblock \doi{10.12942/lrr-2011-7}.
\newblock URL \url{https://doi.org/10.12942/lrr-2011-7}.

\bibitem{fokker1929}
A.~D. Fokker, Ein invarianter variationssatz f\"ur die bewegung mehrerer elektrischer massenteilchen, \emph{Zeitschrift f\"ur Physik}. {\bf 58}, \penalty0 386--393  (1929).

\bibitem{hogarth1962}
J.~E. Hogarth, Cosmological considerations of the absorber theory of radiation, \emph{Proceedings of the Royal Society A}. {\bf 267}, \penalty0 365--383  (1962).

\bibitem{hadamard1923}
J.~Hadamard, \emph{Lectures on Cauchy's Problem in Linear Partial Differential Equations}. Yale University Press, New Haven  (1923).

\bibitem{friedlander1975}
F.~G. Friedlander, \emph{The Wave Equation on a Curved Space-Time}. Cambridge University Press, Cambridge  (1975).

\bibitem{dewittbrehme1960}
B.~S. DeWitt and R.~W. Brehme, Radiation damping in a gravitational field, \emph{Annals of Physics}. {\bf 9}, \penalty0 220--259  (1960).

\bibitem{narlikar1970}
J.~V. Narlikar, On the general correspondence between field theories and the theories of direct interparticle action, \emph{Mathematical Proceedings of the Cambridge Philosophical Society}. {\bf 64}, \penalty0 1071–1079  (1968).

\bibitem{unruh1976}
W.~G. Unruh, Notes on black-hole evaporation, \emph{Physical Review D}. {\bf 14}\penalty0 (4), \penalty0 870--892  (1976).
\newblock \doi{10.1103/PhysRevD.14.870}.

\bibitem{dewitt1979}
B.~S. DeWitt.
\newblock Quantum gravity: The new synthesis.
\newblock In eds. S.~W. Hawking and W.~Israel, \emph{General Relativity: An Einstein Centenary Survey}, pp. 680--745. Cambridge University Press, Cambridge, England  (1979).

\bibitem{vidal2002}
G.~Vidal, J.~I. Latorre, E.~Rico, and A.~Kitaev, Entanglement in quantum critical phenomena, \emph{Phys. Rev. Lett.} {\bf 90}, \penalty0 227902  (Jun, 2003).
\newblock \doi{10.1103/PhysRevLett.90.227902}.
\newblock URL \url{https://link.aps.org/doi/10.1103/PhysRevLett.90.227902}.

\bibitem{reznik2003}
B.~Reznik, Entanglement from the vacuum, \emph{Foundations of Physics}. {\bf 33}\penalty0 (1), \penalty0 167--176  (2003).
\newblock \doi{10.1023/A:1022875910744}.

\bibitem{hari2024}
K.~Hari, S.~Barman, and D.~Kothawala, Universal role of curvature in vacuum entanglement, \emph{Phys. Rev. D}. {\bf 109}, \penalty0 065017  (Mar, 2024).
\newblock \doi{10.1103/PhysRevD.109.065017}.
\newblock URL \url{https://link.aps.org/doi/10.1103/PhysRevD.109.065017}.

\bibitem{birrell1982}
N.~D. Birrell and P.~C.~W. Davies, \emph{Quantum Fields in Curved Space}. Cambridge Monographs on Mathematical Physics, Cambridge University Press, Cambridge  (1982).

\bibitem{dewitt1975}
B.~S. DeWitt, Quantum field theory in curved spacetime, \emph{Physics Reports}. {\bf 19}\penalty0 (6), \penalty0 295--357  (1975).
\newblock \doi{10.1016/0370-1573(75)90051-4}.

\bibitem{mach1893}
E.~Mach, \emph{The Science of Mechanics: A Critical and Historical Exposition of its Principles}. Cambridge Library Collection - Physical Sciences, Cambridge University Press  (2013).

\bibitem{zalaletdinov1992}
R.~M. Zalaletdinov, Averaging out the einstein equations, \emph{General Relativity and Gravitation}. {\bf 24}\penalty0 (10), \penalty0 1015--1031  (1992).
\newblock \doi{10.1007/BF00756944}.

\bibitem{kothawala2023-essay}
D.~Kothawala, Limits of a non-local quantum spacetime, \emph{International Journal of Modern Physics D}. {\bf 32}\penalty0 (14), \penalty0 2342021  (2023).
\newblock \doi{10.1142/S021827182342021X}.

\bibitem{padmanabhan2010entropy}
T.~Padmanabhan, Thermodynamical aspects of gravity: New insights, \emph{Reports on Progress in Physics}. {\bf 73}\penalty0 (4), \penalty0 046901  (2010).
\newblock \doi{10.1088/0034-4885/73/4/046901}.

\bibitem{wainwright2003}
M.~Wainwright, N.~Wickramasinghe, J.~Narlikar, and P.~Rajaratnam, Microorganisms cultured from stratospheric air samples obtained at 41 km, \emph{FEMS Microbiology Letters}. {\bf 218}\penalty0 (1), \penalty0 161--165  (01, 2003).
\newblock ISSN 0378-1097.
\newblock \doi{10.1111/j.1574-6968.2003.tb11513.x}.
\newblock URL \url{https://doi.org/10.1111/j.1574-6968.2003.tb11513.x}.

\end{thebibliography}


\end{document}